\documentclass{IEEEtran}
\usepackage{amsmath,amssymb,amsfonts}
\usepackage{algorithmic}
\usepackage{graphicx}
\usepackage{microtype}
\usepackage{blindtext}
\usepackage{textcomp}
\usepackage{xcolor}
\usepackage{subcaption}
\usepackage{tabularx}
\usepackage{cite}
\usepackage[autostyle]{csquotes}  
\newcolumntype{L}{>{\raggedright\arraybackslash}X}

\usepackage{amsmath}
\DeclareMathOperator{\sgn}{sgn}

\def\BibTeX{{\rm B\kern-.05em{\sc i\kern-.025em b}\kern-.08em
    T\kern-.1667em\lower.7ex\hbox{E}\kern-.125emX}}

\begin{document}

\title{Knowledge-based model validation\\ using a custom metric}

\author{Nicola Henkelmann$^1$ \and Stephan Rhode$^1$  \and Johannes von Keler$^1$ \and \\$^1$Bosch Research, Robert Bosch GmbH,	Renningen, 71272, Germany
\\Email: {{nicola.henkelmann|stephan.rhode|johannes.vonkeler}}@de.bosch.com}
\maketitle

\begin{abstract}
 Vehicle models have a long history of research and as of today are able to model the involved physics in a reasonable manner. However, each new vehicle has its new characteristics or parameters. The identification of these is the main task of an engineer. To validate whether the correct parameter set has been chosen is a tedious task and often can only be performed by experts. Metrics known commonly used in literature are able to compare different results under certain aspects. However, they fail to answer the question: Are the models accurate enough? In this article, we propose the usage of a custom metric trained on the knowledge of experts to tackle this problem.\\
 Our approach involves three main steps: first, the formalized collection of subject matter experts' opinion on the question: Having seen the measurement and simulation time series in comparison, is the model quality sufficient? From this step, we obtain a data set that is able to quantify the sufficiency of a simulation result based on a comparison to corresponding experimental data. In a second step, we compute common model metrics on the measurement and simulation time series and use these model metrics as features to a regression model. Third, we fit a regression model to the experts' opinions. This regression model, i.e., our custom metric, can than predict the sufficiency of a new simulation result and gives a confidence on this prediction.
\end{abstract}

\section{Introduction}

Modern cyber-physical systems are of ever increasing complexity and demand for large experimental campaigns to ensure the safety of a system in operation. For instance, in the context of the development process for autonomous vehicles, estimates suggest decades of necessary experimental testing due to the increased complexity of the system~\cite{Kalra2016,Abdellatif2019}. To alleviate this burden, complementary simulation models are used to predict the system behavior using simulations and partially virtualize the development process \cite{Lutz2017,Abdellatif2019,Riedmaier2021a,Dona2022,Viehof2018,Neurohr2023,Danquah2020}. It is of crucial importance for these computational models to fit to their purpose, i.e., to predict the quantities of interest in areliable and accurate enough manner.\\
The effort and expertise required to provide proof and evidence for this effort can easily be  underestimated, given the complex argumentation chain that must be established. The process begins with identifying the appropriate quantities of interest~(QoI) for the cyber-physical system under investigation, a challenge of itself~(\cite[Chapter~3]{Vincenti1990}). This identification is followed by defining suitable metrics to evaluate the sound functioning of the cyber-physical system it-self \cite{Schultz1961}. Finally, one has to determine an appropriate notion of "accurate enough", i.e., a threshold for this metric in the simulation context. Hence it should be no surprise that there exist a lot of different ideas, approaches and methods how to handle this challenging task.\\
In the context of this triad (QoIs, metric, and threshold) our approach combines the necessary expertise of the subject matter specialist with a broad basis of well established metrics to build a custom metric which provides a use-case specific notion of "accurate enough", while disclosing its inherent uncertainty. \\         
The general question of quantifying the similarity or discrepancy between systems and their representations is at the heart of science and engineering. There are many different view points and methods how to approach this topic (cf. \cite{Oberkampf_Roy_2010} Chapter 2.1 and 10.1).   
In engineering the most common validation methods in general and especially for simulations are based on response analysis. Here a simulation response on given input data is compared with a reference, which is gained from experimental data (cf. \cite{Viehof2018, Lee2019, Riedmaier2021a, Dona2022} and reference therein). The review on validation methods in~\cite{Dona2022} classified response analysis into:
graphical comparisons (also known as face validation) \cite{Sargent2010}; scalar quantities like key performance indices (KPIs)\cite{ObBa2006, Dowding2008, FeObGi2008};
time-histories \cite{Geers1984, Poolla1994, Sarin2008, Mahadevan2017}; frequency analysis \cite{Halder2013}; statistical testing \cite{Rebba2006, Liu2011}; and
statistical discrepancy analysis \cite{Mayer1993, Klein1995, FeObGi2008, Halder2012, Rhode2019}. 

\subsection{Metrics used in validation}

Among response analysis methods, metrics which map two time-series into a measure are a prominent tool. However, the wide variety of available metrics in literature already suggests that choosing "the right one" is a complex task. In the time domain, the most prominent metric is the root mean squared error between two time-series $\mathbf{x}$ and $\mathbf{y}$ of equal size $N$
\begin{equation}
\text{RMSE}=\sqrt{\frac{\sum_{i=1}^N (\mathbf{x}_i - \mathbf{y}_i)^2}{N}}.
\end{equation}

There are extensions of RMSE which allow to compare different time series on varying scales through normalization. Common choices for the normalization are the range or mean of the reference time-series $\mathbf{y}$, for instance $\text{NMRSE} = \text{RMSE} / (\max(\mathbf{y}) - \min(\mathbf{y}))$.

Another popular standard measure is the coefficient of correlation
\begin{equation}
\rho = \frac{\sum\limits_{i=1}^N (\mathbf{x}_i- \overline{\mathbf{x}})(\mathbf{y}_i- \overline{\mathbf{y}})}{\sqrt{\sum\limits_{i=1}^N (\mathbf{x}_i- \overline{\mathbf{x}})^2 \sum\limits_{i=1}^N (\mathbf{y}_i- \overline{\mathbf{y}})^2}},
\end{equation}

which is also known in its square form as the coefficient of determination R-square. The coefficient of correlation indicates the extend of linear relationship between two time-series and the R-square is an often used measure for the goodness of fit. However, RMSE, coefficient of correlation, and R-square are sensitive to phase difference between $\mathbf{x}$ and $\mathbf{y}$, which means that a small phase error, which might occur in a measurement setup, causes variation in these standard measures. A separation into error caused by phase and magnitude requires more sophisticated measures.

Cross correlation 
\begin{equation}
\rho(n) = \sum_{i=-\infty}^{\infty} \mathbf{x}_{i+n} * \mathbf{y}_i
\end{equation}
measures the similarity of two time-series as a function of the displacement of one relative to the other. Cross correlation is used in many combined metrics to determine an optimal shift $n*$ of one signal to another in order to separate phase and magnitude error. The optimal shift is defined by $n$ where cross correlation gives its maximum value. This approach leads to combined metrics, which overcome the problem that a single error measure cannot describe the match between time-series for all cases.

Combined metrics have been proposed by Sprague and Geers~\cite{Geers1984,Sprague2004,Schwer2007}, Russel~\cite{Russell1997}, Sarin et.al.~\cite{Sarin2008}, and have been standardized for dynamical models of road vehicles in~\cite{ISO18571}. The combined metric of Sprague and Geers consists of magnitude and phase error ($M_{\text{SG}}$, $P_{\text{SG}}$), which are computed by

\begin{equation}
\psi_{\text{XX}} = \frac{\sum_{i=1}^N \mathbf{x}_i^2}{N},
\psi_{\text{YY}} = \frac{\sum_{i=1}^N \mathbf{y}_i^2}{N},
\psi_{\text{XY}} = \frac{\sum_{i=1}^N \mathbf{x}_i y_i}{N},
\end{equation}
\begin{equation}
M_{\text{SG}} = \sqrt{\frac{\psi_{\text{XX}}}{\psi_{\text{YY}}}} - 1,
\end{equation}
\begin{equation}
P_{\text{SG}} = \frac{1}{\pi} \cos^{-1} \left(\frac{\psi_{\text{XY}}}{\sqrt{\psi_{\text{XX}}\psi_{\text{YY}}}} \right).
\end{equation}
The combined error is defined as
\begin{equation}
C_{\text{SG}} = \sqrt{M_{\text{SG}}^2 + P_{\text{SG}}^2}.
\end{equation}

Russel~\cite{Russell1997} proposed magnitude phase and a combined error measure, which is similar to the Sprague and Geers metric. However, the magnitude equation 

\begin{equation}
M_{\text{R}} = \sgn (\psi_{\text{XX}} - \psi_{\text{YY}}) \log_{10}\!
\left( 1 + \left| \frac{\psi_{\text{XX}} - \psi_{\text{YY}}}{\sqrt{\psi_{\text{XX}}\psi_{\text{YY}}}} \right| \right)
\end{equation}

is modified to match approximately the same scale as the phase error. Although the separation into magnitude and phase error of Sprague and Geers and Russel's metric allows more fine granular investigation in the source of the model error than a single metric like RMSE, \cite{Sarin2008} found counter examples, where both metrics fail to observe a magnitude error due to missing shape information in both metrics. 

The normalized integral square error (NISE)~\cite{Donnelly1983} was developed to measure the error of repeated time series of side impact tests and considers cross-correlation $\rho$ concept. NISE consists of phase shift, amplitude difference (magnitude error), and shape difference

\begin{equation}
P_{\text{NISE}} = \frac{2 \psi_{\text{XY}} (n*) - 2 \psi_{\text{XY}}}{ \psi_{\text{XX}} + \psi_{\text{YY}}}
\end{equation}
\begin{equation}
M_{\text{NISE}} = \rho(n*) - \frac{2 \psi_{\text{XY}} (n*)}{ \psi_{\text{XX}} + \psi_{\text{YY}}}
\end{equation}
\begin{equation}
S_{\text{NISE}} = 1 - \rho(n*)
\end{equation}
\begin{equation}
C_{\text{NISE}} = P_{\text{NISE}} + M_{\text{NISE}} + S_{\text{NISE}} = 1 - \frac{2 \psi_{\text{XY}}}{\psi_{\text{XX}} + \psi_{\text{YY}}}.
\end{equation}

One of the time series is shifted by $n*$ steps to compensate phase error in the $\psi_{\text{XY}}$ term before the magnitude error is determined. However, the shape error is canceled out in the overall measure and hence, NISE does not account for shape differences between $\mathbf{x}$ and $\mathbf{y}$. 

Consequently, the enhanced error assessment of response time histories (EEARTH) metric in~\cite{Sarin2008} consists of three terms, the phase, magnitude, and slope error which utilizes vector-norms, correlation measures, and the dynamic time warping (DTW) algorithm to measure similarities between time series which may vary in speed. The phase error is determined by a correlation measure and the magnitude error was designed with a global correlation based shift and a local DTW based signal preprocessing step. A clear separation into phase and magnitude error is often a challenge for combined signal metrics. A small phase deviation between two time series causes large magnitude error when standard metrics are applied. Hence, in addition to a global signal shift by $n*$ steps, DTW is used to warp the signal locally to cancel out the phase error coupling in the magnitude error, which is computed by the $L_1$ vector norm on the shifted and warped signals. The slope error is also computed by the $L_1$ vector norm from derivatives of the shifted time series. DTW is also here applied to remove the local coupling of phase error and slope error. The combined metric 
\begin{equation}
\text{EEARTH} = 10 - (w_1 P_{\text{EEARTH}} + w_2 M_{\text{EEARTH}} + w_1 S_{\text{EEARTH}} )
\end{equation}
is built from a weighted sum of phase, magnitude, and slope error. As proposed by Sarin et al.~\cite{Sarin2008}, the weights ($w_1 \dots w_3$) were found by linear regression on 15 face validation plots, which were labeled individually by six subject matter experts. The ratings ranged form 1 (worst) to 10 (excellent) and 10 plots rated by the experts were used for training of the metric, whereas 5 plots were used for validation. This procedure of expert guided metric design is one rare reference of a custom metric and our approach extends the idea in~\cite{Sarin2008} by machine learning methods from feature engineering and feature selection. Compared with previous single and combined metrics, EEARTH encodes expert knowledge for a specific use case and allows validation tools which guide and learn from traditional face validation.
The most elaborate combined metric to our knowledge is the objective rating metric for non-ambiguous signal for road vehicles, standardized in ISO18571~\cite{ISO18571}. Additional to phase, magnitude, and slope score, ISO18671 considers a corridor score, which rates the signal under test within an inner and outer corridor around the reference signal. The combined metric is similar to the previous combined metrics a weighted sum of all scores and becomes
\begin{equation}
R_{\text{ISO}} = 0.4 \cdot Z_{\text{ISO}} + 0.2 \cdot P_{\text{ISO}} + 0.2 \cdot M_{\text{ISO}} + 0.2 \cdot S_{\text{ISO}},
\end{equation}
where $Z_{\text{ISO}} \dots S_{\text{ISO}}$ denotes corridor, phase, magnitude, and slope score. Note that the weights in the ISO18571 metric are fixed and there is no detailed explanation how they were adjusted. The ISO18571 metric applies also DTW as preprocessing step in the magnitude score and considers a shift correction in the magnitude and slope score. The overall rating $R_{\text{ISO}}$ is interpreted in model grade and ranking, which is shown in Table~\ref{tab:expertrating}. Fig.~\ref{fig:ratingbar}, further disucssed in section~\ref{sec:facevalidation}, provides a colorbar for this rating and shows that the model rating increases progressively. The rating for a fair model is within $0.58 < R \leq 0.8$, whereas an excellent model's rating ranges above $R > 0.94$. \\
Based on these commonly used weighted combined metrics, some recent advance has been made in optimizing the metrics by adjusting their weights~\cite{Widner2022} to specific use case scenarios.

\subsection{Common issues in metric choice}\label{sec:common_issues}
\begin{figure}[h]
	\centering
	\begin{subfigure}{0.4\textwidth}
		\includegraphics[width=\textwidth]{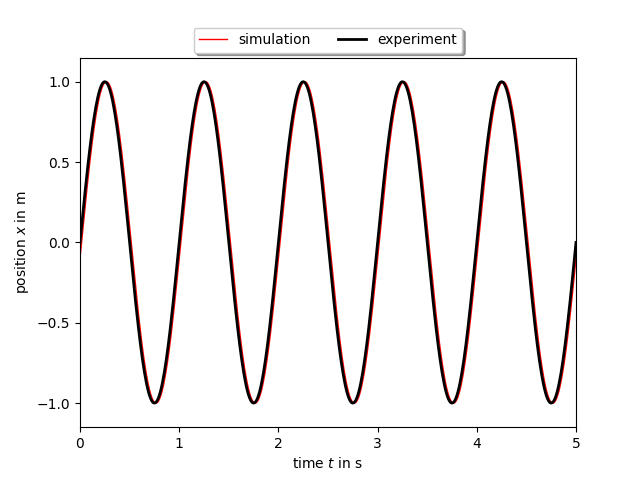}
		\caption{Sine with different phase in simulation}
		\label{fig:phase}
	\end{subfigure}
	\hfill
	\begin{subfigure}{0.4\textwidth}
		\includegraphics[width=\textwidth]{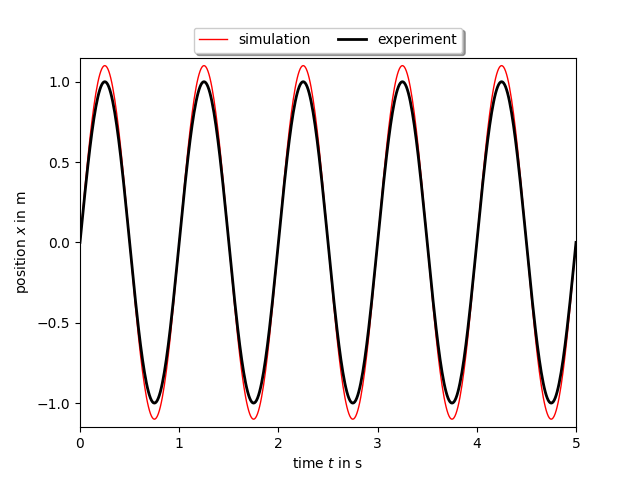}
		\caption{Sine with different amplitude in simulation}
		\label{fig:ampl}
	\end{subfigure}
	\caption{Comparison of sine signals with same RMS-error value}
	\label{fig:phaseampl}
\end{figure}

The many available metrics in literature suggest that it is often not clear which metric is suitable for the specific use case. To underline this statement, some drawbacks of the highly popular (normalized) root mean square metric are demonstrated in the two following examples: In Fig.~\ref{fig:phaseampl}, two sine functions are compared in each plot. We assume that the first sine excitation corresponds to a fictional position of an experiment in black. This fictional simulated is modeled two different models. The fictional prediction of these models is shown in red. In Fig.~\ref{fig:phase}, simulation and experiment differ in a phase shift of $0.1$s. The visibly apparent difference in both signals is quite small. Most readers might agree that in many applications the deviation between simulation and experimental results in this case of a slight phase shift is still acceptable. The root mean square error of the respective curves is $5e-3$mm. On the other hand, in Fig.~\ref{fig:ampl}, the phase of both curves match but the amplitudes differ. Here, the deviation between the two curves is visibly prominent and the simulation accuracy might not be sufficient for many use cases. Yet, the root mean square error here is exactly the same as for the phase shift example: $5e-3$mm. This illustrates the issue that the widely common root mean square error is highly sensitive to a phase shift in two data curves. This might be desired for some use cases, but for others it is not. A good metric is custom to its use case and it is often not sufficient to rely on the most popular known metrics in literature.

This example illustrates common problems in the "right" choice of a metric. By proposing the use of a custom metric and in the article at hand, we want to offer a systematic approach to overcome these issues and find a metric that reflects the uniqueness of each use case. 

\subsection{Uncertainty in validation}
All up to now presented metrics do not consider uncertainty or probabilistic effects which appear in repeated measurements or stem from uncertainty in model parameters, which require additional uncertainty quantification methods to consider for instance measurement noise. Hence, recent developments also include the prediction of uncertainty in the metric~\cite{Xi2015,Xu2016}. Another approach is to apply a statistical based validation framework like the p-box method \cite{Oberkampf_Roy_2010, ROY2011, Riedmaier2021} for scalar quantities of interest, statistical validation based on distribution divergence \cite{Rhode2019}, or methods based on confidence bands \cite{Rhode2020}.\\
The main issue in choosing a suitable validation metric is that the source of error between experimental and simulation data is unknown. The deviation may be due to measurement noise, parametric uncertainties, the model abstraction level or uncertainties in the excitation~\cite[chapter~9]{Bendersen2011},\cite[chapter~8]{nasa2010}. To be more precise, a suitable metric is highly dependent on the measurement device, the system and the simulation model. This knowledge has lead to the wide usage of subjective validation techniques~\cite{Viehof2017} in the daily engineering practice. 

\subsection{Outline}
The paper is structured as follows: After the introduction of known metrics and their limitations upon here, in section~\ref{sec:common_issues}, we discuss some common issues in general purpose metrics to motivate the development of a custom metric. Next, the proposed method is described in detail in section~\ref{sec:method}.\\
Subsequently, we demonstrate its application to three different use cases in section~\ref{sec:application}. The first use case in section~\ref{sec:application_art} is an artificially setup example. It serves to generate test data under different testing conditions and investigate the dependencies in the method in detail. After gaining this insights, we apply the method to two real world use cases in a second step. One is taken from literature and uses data of radial velocities in a geological medium, cf.~section~\ref{sec:lit_data}. The second use case is a rack position controller in a steering system, cf.~\ref{sec:rpc}.\\
Section~~\ref{sec:conclusion} gives a short summary and outlook.

\section{Method}\label{sec:method}

The main idea of the proposed custom metric is to collect expert ratings and subsequently find a combination of metrics that adequately describe them. The framework can be viewed as a machine learning approach that translates the expert opinion (data) into a parametric surrogate model. In the following we first describe the data collection process in section~\ref{sec:facevalidation}. As the evaluation of whether a curve "fits" to another is subjective, the main focus here is to find a suitable rating process targeted to provide clear reference for the experts in the labeling process. Based on that data, the custom metric is extracted in section~\ref{sec:custommetric}. This is done by fitting a linear regression model to a set of pre-selected metrics on the test data to the expert ratings. In addition, the computation of a confidence interval on the model predictions is shown in section~\ref{sec:custommetric}. This computation is crucial step as the metric is designed to be used in validation.

\subsection{Face validation}\label{sec:facevalidation}

\begin{figure}
	\centering
	\includegraphics[width=0.48\textwidth]{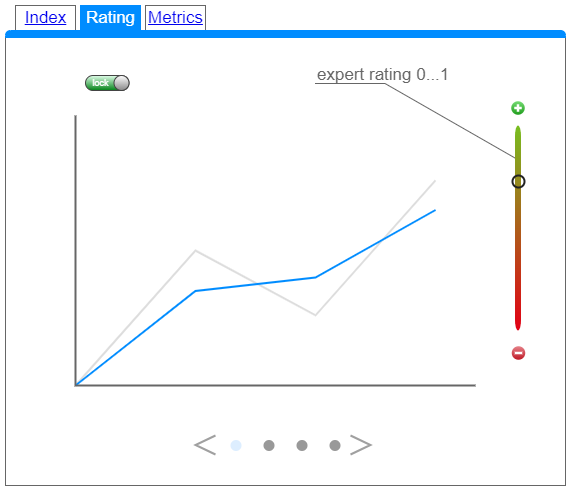}
	\caption{Interactive rating document. The graph shows timeseries data of simulation and measurement next to the rating bar.}
	\label{fig:rating_doc}
\end{figure}

Face validation is widely used in engineering to validate a simulation model~\cite{Carson2022,Harris2020,Chapurlat2010}. It refers to the process of subject matter experts looking at the simulation outputs as well as in some cases additional respective measurement data and evaluating the validity of the model based on that data and their previous knowledge. In the work at hand, we assume that simulation output and measurement data is given and the face validation happens through a comparison of both data sets.\\
The evaluation of "fit" or "no fit" is intuitively often framed as a classification task. Asked about his opinion, an expert will most certainly answer in terms of a description (ok, good, perfect, not at all). However, for statistical analysis of the data it is highly advantageous to have a continuous scale of rating. Thus, we asked the experts to rate the fit on a continuous scale from 0 to 1. To ensure the same understanding of the significance of, e.g., the rating 0.7, we propose to provide the experts with the descriptions of each value range in Tab.~\ref{tab:expertrating}, which was adopted from \cite{ISO18571}. This table was given as a primary explanation to each of the experts at the top of an interactive document shown in Fig.~\ref{fig:rating_doc}. The first column defines the rank $r$ in integer values, to which a descriptive grade in natural language (excellent, good, fair, poor) is attached in the second column. To translate this classification to a continuous scale, for each of the previously defined categories a rating $R$ is given in the third column. The rating defines a range of continuous values belonging to the categories. To complete the explanation, a more detailed description of deviations in each category is given in a fourth column.

\begin{table}
	\centering
	\caption{{Rating of model quality in accordance with \cite{ISO18571}.}}
	\begin{tabularx}{\linewidth}{c|c|c|L}
		\textbf{Rank} $r$& \textbf{Grade} & \textbf{Rating} $R$ & \textbf{Description}\\\hline \hline
		1 & Excellent & $R > 0.94$ & Almost perfect characteristics \newline of the reference signal is captured \\\hline
		2 & Good & $0.8 < R \leq 0.94$ & Reasonably good characteristics of the reference signal is captured, but there are noticeable differences between them\\\hline
		3 & Fair & $0.58 < R \leq 0.8$ & Basic characteristics of the reference signal is captured but there are significant differences between the signals\\\hline
		4 & Poor & $R \leq 0.58$ & Almost no correlation between the two signals
	\end{tabularx}
	\label{tab:expertrating}
\end{table}

\begin{figure}
	\centering
	\includegraphics[width=0.48\textwidth, trim=50 0 50 0, clip ]{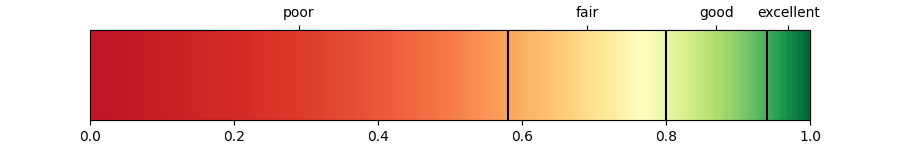}
	\caption{Colorbar for expert rating.}
	\label{fig:ratingbar}
\end{figure}

With this explanation, the experts further find the rating bar shown in Fig.~\ref{fig:ratingbar} attached to an interactive slider next to each of the labeling scenarios in the interactive rating document. Their label is created by their adjustment of the slider on the rating bar.

Whether the prediction of a simulation model fits to a measurement time series of a system can depend on various factors. These factors include the model itself, which, due to a certain level of abstraction, does not account for some of the physical effects visible in the data set. Typical examples are linearizations of more complex models due to computational costs or simplified models of complex physical effects (e.g., friction coefficients). Additionally, the evaluation may strongly depend on the use case, i.e., the same model may be sufficient for one use case while it lacks the necessary accuracy for another. For example, if the specification of a steering system only requires the car to stay with a predefined band around a lane, larger errors in the dynamic model are acceptable. Dependent on which of this information is available to the experts in the labeling process, the domain of applicablability for the derived custom metric is different. I.e., if fewer information is given to the experts a wider application domain is targeted. This may seem as the desirable choice but certainly comes with a trade-off in accuracy.

\subsection{Custom metric}\label{sec:custommetric}

\begin{figure}[h]
	\centering
	\includegraphics[width=0.6\columnwidth]{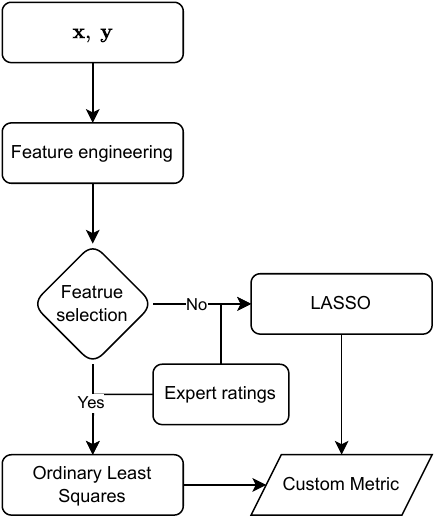}
	\caption{Steps to design a custom metric.}
	\label{fig:process}
\end{figure}

Figure~\ref{fig:process} shows the steps to derive the custom metric. We consider two ways, one with feature selection and conventional linear regression to derive the custom metric and the other through least absolute shrinkage and selection operator (LASSO)~\cite{Tibshirani1996}, which variable selection and regression in one step.

In contrast to existing literature in validation metric design, we combine model selection and parameter estimation in the custom metric. Metric selection means to pick a metric model from a list of possible models, which aim to model expert ratings of face validation plots. Conventionally, the presented combined metrics like Sprague and Geers~\cite{Geers1984,Sprague2004,Schwer2007}, or ISO18571~\cite{ISO18571} cover two to four model terms like magnitude, phase, slope, and corridor error to model the overall rating of a validation task. In the custom metric, we gather expert ratings from independent face validation plots and create the model terms from a list of basic metrics which we implemented in a metrics library. Hence, the general custom metric is a linear combination of ansatz functions called features $f$ in the sequel, which are weighted by $w$ and a residual term $\epsilon$ is added to account for uncertainty in the custom metric rating $R$

\begin{equation}
R = \sum\limits_{i=1}^N \mathbf{w}_i \mathbf{f_i(\mathbf{x}, \mathbf{y})} + \epsilon.
\end{equation}

The weights are adjusted by linear regression or the more advanced robust regression like least trimmed squares \cite{Rousseeuw2005} in case of outliers in the data. This process of model creation, model selection, and parameter estimation from a set of ansatz models is also known as symbolic regression \cite{LaCava2021} and is applied to numerous engineering problems for instance in automatic materials modeling in \cite{Flaschel2021}. In this study, a start model with numerous terms of ansatz models was fitted to data through regularized linear regression to obtain a sparse model with optimized model parameters in one shot. \\
The model fitting is done on a subset of the available labeled data set, the training data. The remaining data is excluded from the model training and later used as a test data set. We are interested in the uncertainty of the prediction. For linear regression under the assumption of normally distributed weights, the prediction interval is given by~\cite{Fahrmeier2016}
\begin{align*}
	\hat{w} \pm \hat{\sigma}_\text{train} t_{1-\alpha/2}(n-2)
\end{align*}
where $n$ is the sample size and $t(n)$ the student's t-distribution. The sample variance $\hat{\sigma}_\text{train}$ is given by
\begin{align*}
	\hat{\sigma}_\text{train}^2 = \frac{1}{n-p-1} \sum_{i=1}^N (Y_i - \hat{Y}_i)^2.
\end{align*}
Here, $p$ is the number of parameters (or coefficients in regression) used in the model. For a sample size $n>30$, the student's t-distribution can be approximately replaced with the normal distribution
\begin{align}\label{eq:conf_1d}
	\hat{w} \pm \hat{\sigma}_\text{train} q_{\mathcal{N}} (1-\alpha/2).
\end{align}
In the following, the exact formulation will be used if not stated otherwise. Here, multi-dimensional dependencies of the weights are neglected. For multidimensional linear regression, the prediction interval is given as 
\begin{align}\label{eq:conf_multi}
	\hat{w}_\text{multi} \pm \hat{\sigma}_\text{multi}  t_{1-\alpha/2}(n-p-1)
\end{align}
with 
\begin{align*}
\hat{\sigma}_\text{multi} = \sqrt{\hat{\sigma}_\text{train} (1+x^T(X^T X) x)}
\end{align*}
\cite[p.~463]{Fahrmeier2016}.

The evaluation of a prediction interval on metrics used for validation is, to the best of the authors' knowledge, something that is missing in literature. This is, however, a crucial step in obtaining a reliable model. In section~\ref{sec:lit_data}, we will demonstrate this on literature data, how a good nominal fit of the data can otherwise lead to false confidence in the custom metric.

\section{Application}\label{sec:application}

We demonstrate the method proposed in section~\ref{sec:method} using two different data sets in the following section. The first data set consists of artificially generated data that enables insights into the underlying sources of uncertainty and its quantification. The second and third data sets are real world example to demonstrate the method in engineering practice. The second one described in section~\ref{sec:lit_data} is taken from literature~\cite{Schwer2007}, the third one in section~\ref{sec:rpc} is using data available to the authors.

\subsection{Application to artificial data}\label{sec:application_art}

To study the method in depth, we apply the method of manufactured universe \cite{Stripling2011}, which was designed to explore statistical and modeling assumptions in validation and uncertainty quantification methods. In short, ``experimental'' data with imperfections are created in a manufactured reality and compared with imperfect model simulations to examine the validation concept under investigation.

\begin{figure}[h]
	\centering
	\includegraphics[width=0.95\columnwidth]{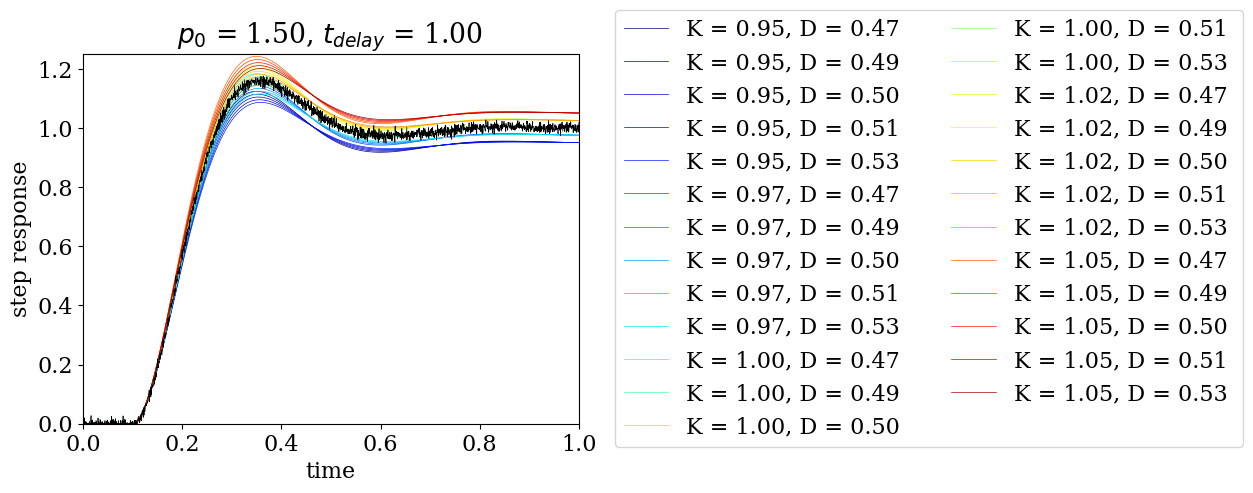}
	\caption{Experimental and simulation data}
	\label{fig:data}
\end{figure}

\begin{figure}[h]
	\centering
	\includegraphics[width=0.95\columnwidth]{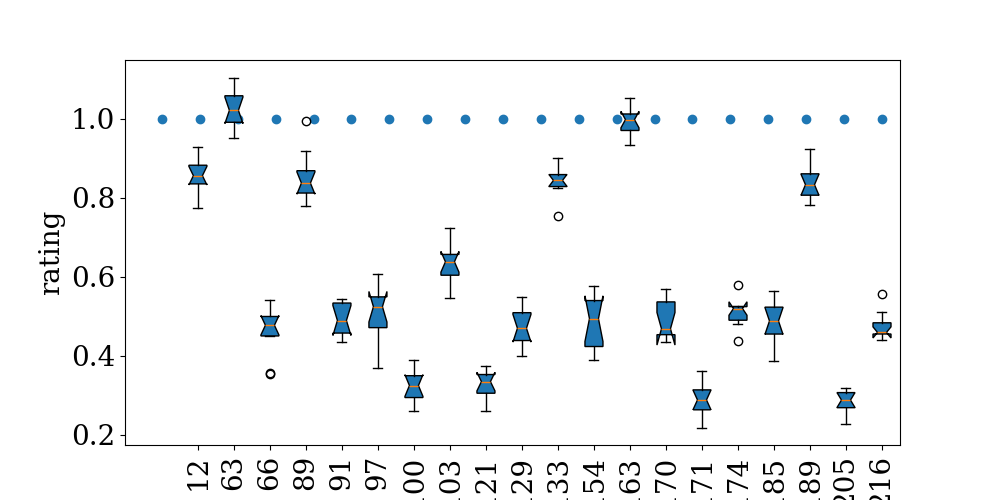}
	\caption{Expert ratings of synthetic data}
	\label{fig:exp_rating_art}
\end{figure}

\begin{figure}[h]
	\centering
	\includegraphics[width=0.95\columnwidth]{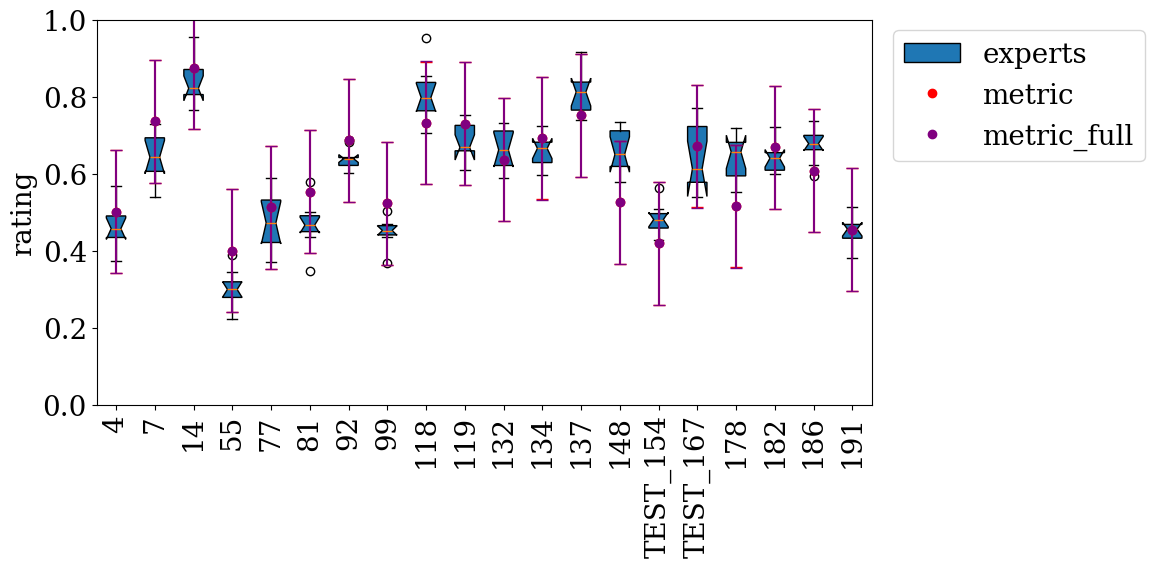}
	\caption{Metric result for synthetic data}
	\label{fig:metric_art}
\end{figure}

\begin{figure}[h]
	\centering
	\includegraphics[width=0.95\columnwidth]{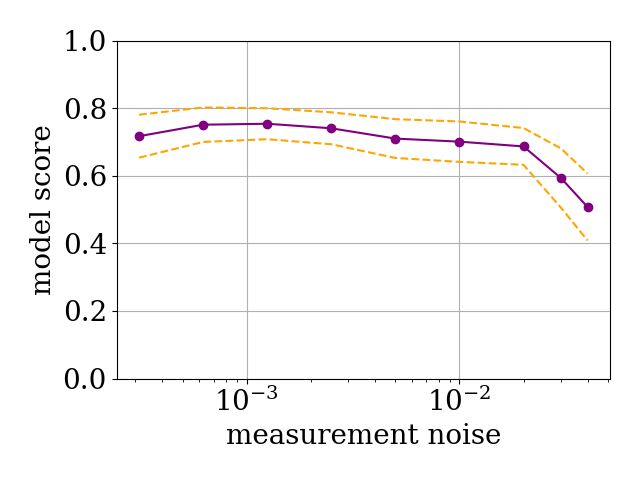}
	\caption{Model score dependent on measurement noise}
	\label{fig:score_measnoise}
\end{figure}

\begin{figure}[h]
	\centering
	\includegraphics[width=0.95\columnwidth]{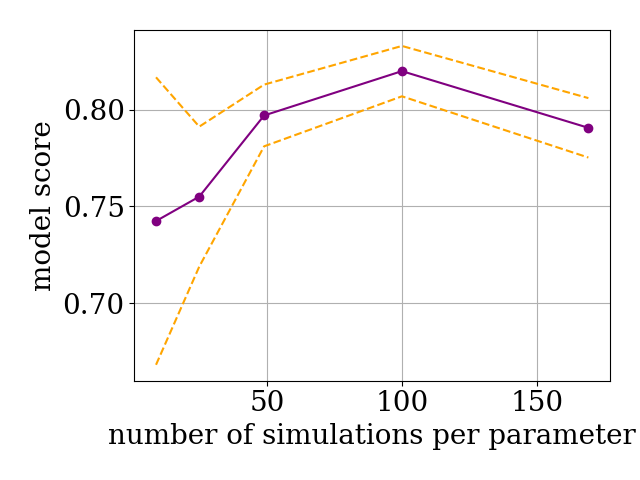}
	\caption{Model score dependent on number of simulations}
	\label{fig:score_numsim}
\end{figure}

\begin{figure}[h]
	\centering
	\includegraphics[width=0.95\columnwidth]{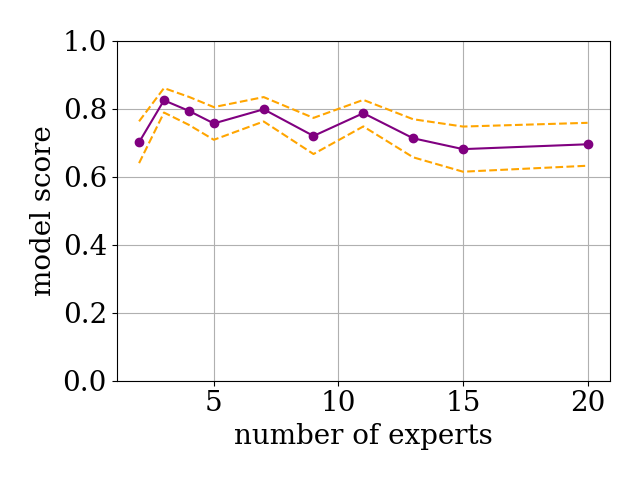}
	\caption{Model score dependent on number of experts}
	\label{fig:score_numexp}
\end{figure}

\begin{figure}[h]
	\centering
	\includegraphics[width=0.95\columnwidth]{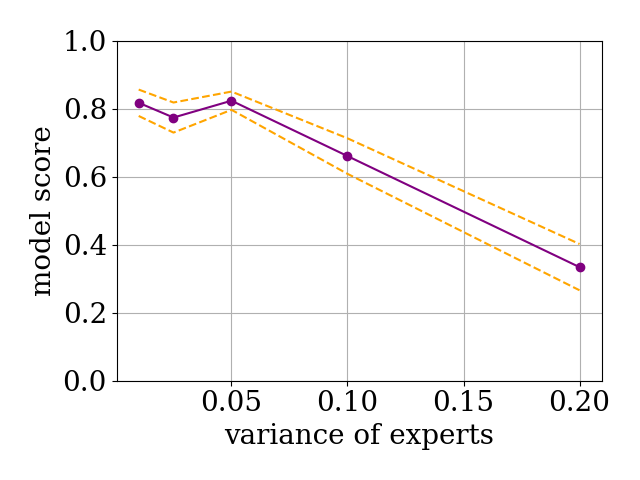}
	\caption{Model score dependent on variance of experts' opinions}
	\label{fig:score_varexp}
\end{figure}

\begin{figure}[h]
	\centering
	\includegraphics[width=0.95\columnwidth]{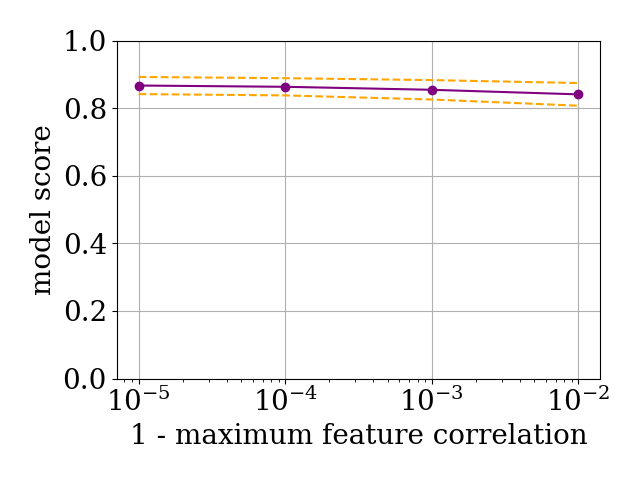}
	\caption{Model score dependent on number of features}
	\label{fig:score_dropcorr}
\end{figure}

We start by generating an artificial test data set from which we can investigate the influence of different parameters, e.g., the influence of measurement noise on the metric quality. As the data set is artificial and not connected to any real use case, we can not perform the face validation step (section~\ref{sec:facevalidation}) for the artificial data but rather rely on artificially generated expert labels. Thus, what is investigated in this study is the derivation and quality of the custom metric itself (section~\ref{sec:custommetric}).\\
We use the step response of a PT2-element as a reference example. The differential equation of the system reads
\begin{align*}
	\dot{\mathbf{x}} = \left( \begin{array}{cc}
	\mathcal{P} & 1  \\
	- \frac{1}{T_0^2} & -\frac{2}{T_0} D_0 \end{array} \right) \mathbf{x} \, + \,
	\left( \begin{array}{c}
	0   \\
	\frac{K_0}{T_0^2}  \end{array} \right) u(t)
\end{align*}
with the reference parameters $T_0$, $K_0$ and $D_0$ and the step excitation
\begin{align*}
	u(t) = \begin{cases}
	1.0 & \text{if } t\geq (0.1+t_\text{delay})\\
	0,              & \text{otherwise.}
	\end{cases}
\end{align*}
We assume that the experimental data is disturbed by a non-zero process noise $\mathcal{P}$ and a measurement noise $\mathcal{M}$
\begin{align*}
	\mathbf{x}_\text{exp} = \mathbf{x} + \mathcal{M}
\end{align*}
where the process noise follows a uniform distribution
\begin{align*}
	\mathcal{P} = \mathcal{U}(-p_0,p_0)
\end{align*}
with the interval radius $p_0$ and the measurement is assumed to follow a Gaussian distribution with zero mean
\begin{align*}
	\mathcal{M} = \mathcal{N}(0,\sigma_\text{measurement}).
\end{align*}
Additionally, a time delay in the measurement data is added. This is done by adding a $t_\text{delay}> 0.0$ (see Tab.~\ref{tab:ref_param}) for the generation of the experimental data. The artificial experimental data is generated using the reference parameters $D$.\\
The obtained experimental data is compared to simulation data with zero measurement and process noise, as well as zero time delay $t_\text{delay}^\text{sim}=0$. We want to train a metric on how well the parameters in the simulation match the experimental data, i.e., how well the system is identified. Thus, we generate simulation data for a range of parameters
\begin{align*}
	K &= K_0 \times \{0.95 ... 1.15\}\\
	D &= D_0 \times \{0.95 ... 1.15\}.
\end{align*}
The artificial expert ratings $R$ are generated by the relative difference in the parameter values of experiment and simulation
\begin{align}\label{eq:expert}
	R = & 1 - w_\text{K} \| \frac{2(K - K_0)}{K_0+K}  \| + w_\text{D} \| \frac{2(D - D_0)}{D_0+D}  \| \\
	 & + \mathcal{N}_\text{exp} (0,\sigma_\text{exp}),
\end{align}
where $\mathcal{N}_\text{exp}$ is a probabilistic term introduced to represent the subjective evaluation of the metric by different domain experts.\\
We generate a reference data for initial values $\mathbf{x}(0) = [0, 0]^T$ over a time span of $t \in [0, 1]$ using the parameters listed in Tab.~\ref{tab:ref_param}. For the simulation data, we vary $D$ and $K$ within an interval of $[0.95 ... 1.15]$ of the reference values, using an number of relative values distributed equidistantly in the respective interval. For the reference, five values are used; i.e., the total number of simulations is $5 \times 5 \times 3 \times = 225$ (full-factorial for $K$, $D$, $p_0$ and $t_\text{delay}$).
\begin{table}
	\centering
	\caption{Parameters for reference data}\label{tab:ref_param}
	\begin{tabular}{c|c|c}
		Parameter & value  & unit \\ \hline
		$K_0$ & 1.0 & $s$ \\
		$D_0$ & 0.5 & $-$ \\
		$T_0$ & 0.07 & $s$ \\
		$\sigma_\text{measurement}$ & 0.01 & $-$ \\
		$p_0$ & $\{0.0,1.5,3.0\}$ & $1/s$ \\
		$t_\text{delay}$ & $\{0.0,0.5,1.0\}$ & $ms$ \\
	\end{tabular}
\end{table}\\
The resulting data set consists of nine experimental data sets with 25 corresponding simulations. One of the experiments and its simulation data is plotted in Fig.~\ref{fig:data}. Here, the process noise parameter $p_0$ is $1.5$ and the time delay $t_\text{delay}=1$ ms.
\\
For each pairing of experimental and simulation data, ten expert ratings are produced according to equation~\eqref{eq:expert}. The standard deviation in the expert's evaluation is set to $0.05$ and the rating weights to $w_\text{K}=0.7$ and $w_\text{D}=0.7$. The generated expert ratings are shown in Fig.~\ref{fig:exp_rating_art} for a subset of the total data set.
\\
To train the custom metric, we generate a list of metrics for each pair of experimental and simulation data. The used metrics in our case are
\begin{itemize}
	\item mean absolute error
	\item mean squared error
	\item median absolute error
	\item maximum absolute error
	\item coefficient of determination
	\item fraction of explained absolute error
	\item explained variance 
	\item cross correlation
	\item Pearson correlation coefficient
	\item normalized root mean square error
	\item Sprague-Geers amplitude error
	\item Sprague-Geers phase error
	\item Sprague-Geers combined error
	\item ISO18571 metric
	\item corridor score
	\item EEARTH score.
\end{itemize}
Next, we drop correlated features by a correlation threshold of $0.9$. Subsequently, the data is split into $80 \%$ training data and $20 \%$ test data. We than use linear regression and the training data to fit our metric model. The results of the model are shown in Fig.~\ref{fig:metric_art}. The metric is able to capture the expert's evaluations well, both on the training and the test data. The prediction intervals are plotted for a confidence level of $95\%$, both using the naive formula~\eqref{eq:conf_1d} (in red) and the multidimensional formula~\eqref{eq:conf_multi} (in yellow). The remaining descriptive metrics used in the custom metric are mean absolute error, magnitude error and phase error of the Sprague Geers metric, as well as the ISO18571 metric and its corridor score. For this dataset with a high number of training data, the two prediction intervals almost coincide.
\\
In the following, we study the influence of different hyper parameters on the model quality in more depth. We therefor use the model score $m$, defined as the coefficient of determination of the test data on the model. First, the dependence of the model score on the measurement noise $\sigma_\text{measurement}$ in the data set is shown in Fig.~\ref{fig:score_measnoise}. The data is split into test and training data by a random split-selector the model fit to the training data. This randomized process is repeated 50 times, from which a mean score value (purple) and a variance (orange) is extracted and plotted. Fig.~\ref{fig:score_measnoise} shows, that the model quality does not decrease with increasing measurement noise up to very high measurement noise (about 0.05). This means that if the experts exclude measurement noise from their rating, as they do within this artificial scenario, the custom metric excludes it, too. The model successfully fits the metric to the expert's evaluations independent of the measurement noise.
\\
Secondly, we investigate the model score on the number of simulations rated by the experts in Fig.~\ref{fig:score_numsim}. Here we observe that the mean model quality is only slightly affected by the number of simulations in the data. However, if a too small number of simulations is used the variance of the score increases, i.e., the model score becomes more sensitive on the chosen test-training split. In an ideal setting, the average score as well as the variance would stay constant plotted over the numerber of parameters. Howerver, in small data sets, relevant data are more likely to be excluded from the training by a random split strategy. Thus, if a limited number of data is available, the choice of the test-train split is crucial for the model quality. This is a common issue in real data sets, as often the number of measurements is quite limited, see sec.~\ref{sec:lit_data} and sec.~\ref{sec:rpc}.
\\
Next, we are interested in the dependence of the custom metric score on the number of available expert ratings. In Fig.~\ref{fig:score_numexp}, the dependence of the metric quality on the number of experts is shown. The model quality for this example and a Gaussian distribution of the experts opinions does not depend much on the number of experts. The model score is comparable for $2$ or $15$ experts. In reality, having a group of more than $5$ experts available is rather unlikely. For this rather well-behaved artificial example data set, already a low number of experts seems reasonable to obtain a stable model quality.
\\
Additionally, in Fig.~\ref{fig:score_varexp}, the influence of the variance in experts' opinions on the metric is plotted. The variance in the experts' opinions is increased by increasing $\sigma_\text{exp}$ in equation~\eqref{eq:expert}. As expected, the model score of the custom metric decreases with increasing variance. A variance of $0.2$ means, that $50\%$ of the data has a larger distance than $2 \sqrt{2} \approx 0.28$ to the metric mean, i.e, it is not surprising that the custom metric scores poorly for such a large scattering.
\\
Finally, the dependence of the model score on the number of chosen features is investigated in Fig.~\ref{fig:score_dropcorr}. Here, all features with a correlation higher than a maximum feature correlation are dropped. The influence on the model quality is marginal, i.e., the simpler model with less features is able to capture the expert's opinion.
\\
Using this crafted example, we could demonstrate the ability of the custom metric to reflect experts' opinions that typically focuses on system dependent parametric effects, e.g., how well the effect of $K$ is captured in the simulation, rather than common metrics like mean squared error. The custom metric procedure is able to translate signal metrics into expert ratings.

\subsection{Application to literature data}\label{sec:lit_data}

\begin{figure}
	\centering
	\includegraphics[width=0.95\columnwidth]{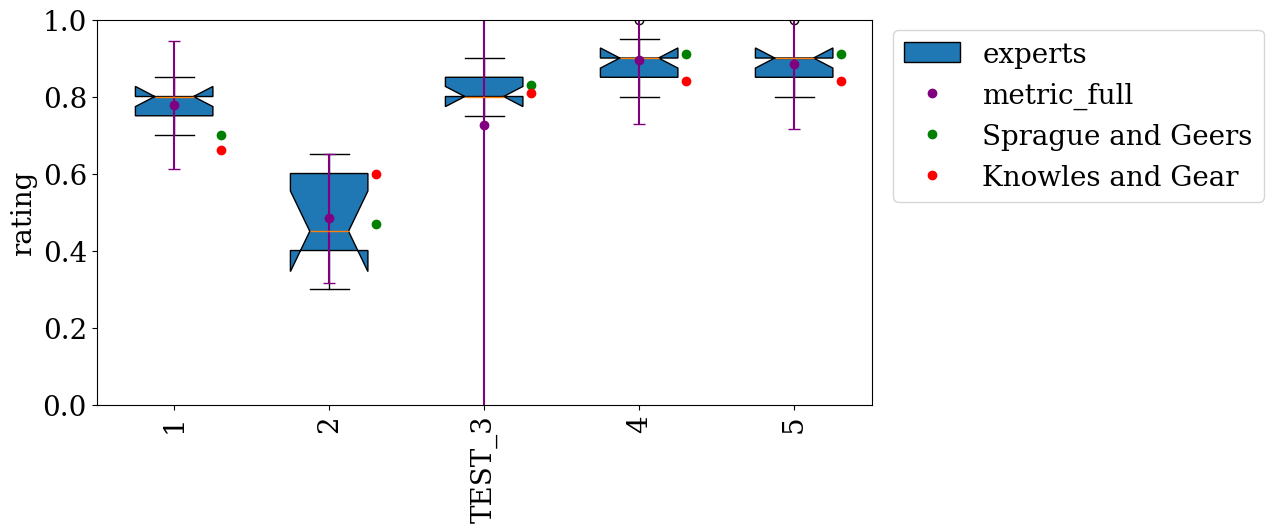}
	\caption{Metric result for literature data}
	\label{fig:metric_lit}
\end{figure}

After studying the custom metric on artificial data in Sec.~\ref{sec:application_art}, we proceed to demonstrate the method on a real world example from the literature~\cite{Schwer2007} in this section. Schwer at a.~\cite{Schwer2007} compared experimental and simulation data of radial velocities in a geological medium due to a nearby energetic source. Eleven subject matter experts have been asked to rate the agreement of experimental data and simulation on a scale from 0 (= very poor) to 1 (=very good). The authors did not further formalize the survey using, e.g., descriptions as listed in Tab.~\ref{tab:expertrating}. The expert ratings for five different time series pairings are plotted in Fig.~\ref{fig:metric_lit}. A significant scattering can be seen, especially as the matching decreases. For pairing number $2$, one expert gave a $0.2$ rating, while another rated it $0.6$. 
The authors compare the Sprague and Geers~\cite{Sprague2003} as well as the Knowles and Gear metric to the subject matter experts' ratings. These metric values are shown as purple and green dots in Fig.~\ref{fig:metric_lit}. They meet the subject matter experts' ratings fairly well. However, as the authors themselves stress in their paper, these validation metrics do not translate to an answer on the \textit{acceptability} of the simulation results: \enquote{[], we do not believe the quantitative value of any metric is important in an absolute sense. [...] It only measures the agreement between computational result and experimental in such a way that positive and negative errors cannot cancel.}~\cite{Oberkampf2002}. Thus, their values have no meaning in terms of a rating but are rater just a metric on a scale from $0$ to $1$ that makes different fits \textit{intercomparable}. Nevertheless, they show some agreement with the experts' ratings for this particular case. In contrast to that, the custom metric shown in yellow in Fig.~\ref{fig:metric_lit} is designed to represent the experts' ratings for this particular domain example. As the data set is rather small, the fitted metric strongly depends on the chosen training and test data set. E.g., sample $2$ is the only one included in the data set with a rating below $0.6$. If this sample is excluded from the training data, the algorithm is not able to learn the metric at lower expert ratings. For the shown test-train-split, i.e., using sample $3$ for testing, the test score is acceptable. For this split, the remaining descriptive metrics in the custom metric are mean absolute error, fraction of explained absolute error, combined error and magnitude of the Sprague-Geers metric as well as the EEARTH score of ISO18751. To obtain a reliable custom metric for the agreement of the radial velocities, a larger data set would be needed. This fact is further emphasized by the rather large prediction intervals of the metric. Yet, we could still demonstrate that the custom metric is able to translate common metrics on the time series data to a custom metric, i.e., a likely rating of subject matter experts.

\subsection{Application to rack position data}\label{sec:rpc}

\begin{figure}
	\centering
	\includegraphics[width=0.95\columnwidth]{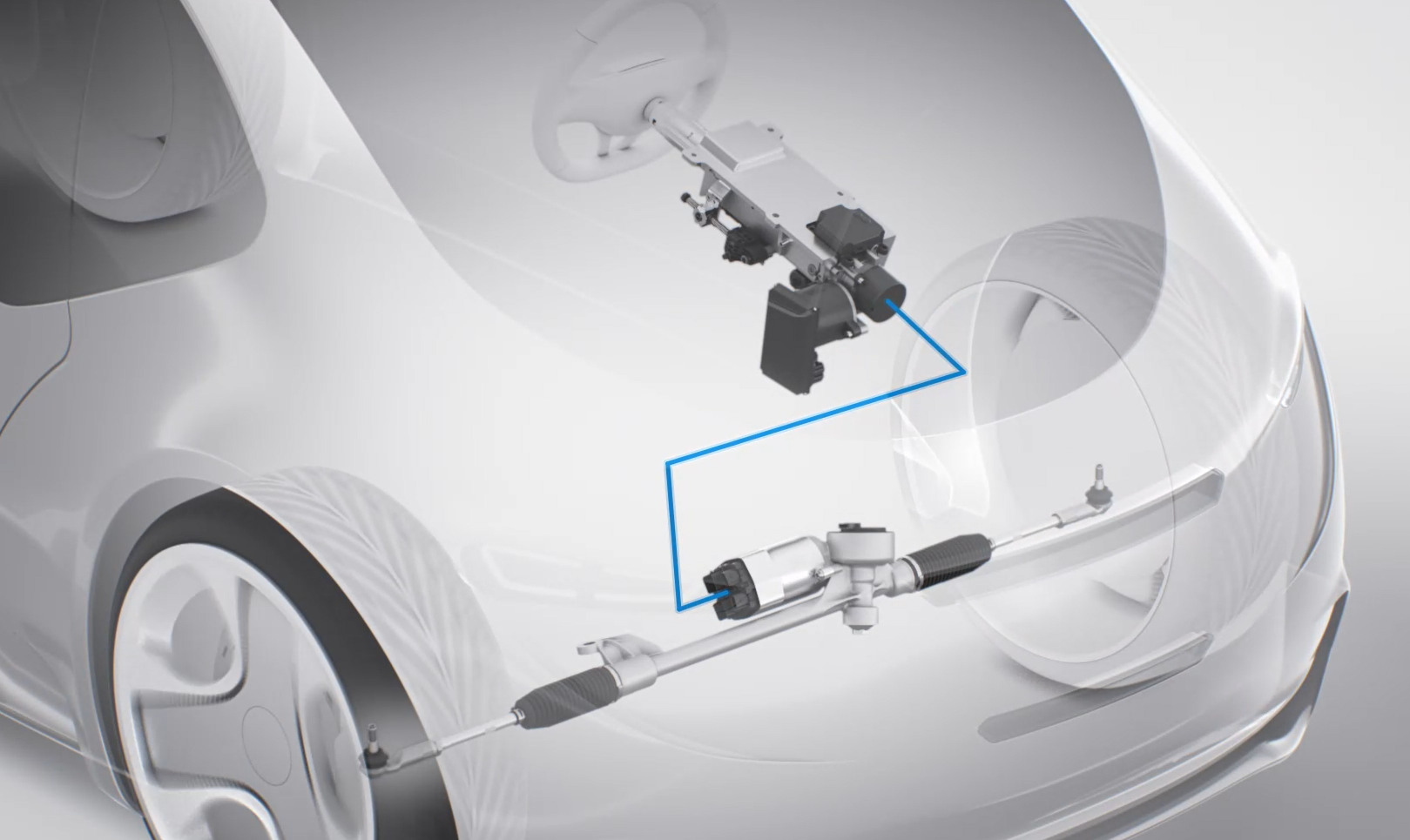}
	\caption{Steer-by-wire system~\cite{AS2024}.}
	\label{fig:steering}
\end{figure}

\begin{figure}
	\centering
	\begin{subfigure}[b]{0.24\textwidth}
		\centering
		\includegraphics[width=0.95\columnwidth]{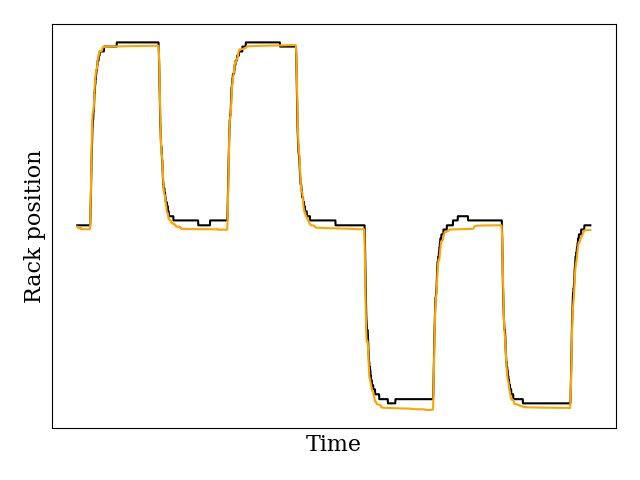}
		\caption{Data for ramp excitation}\label{fig:sim_meas_data_rpc1}
	\end{subfigure}
	\hfill
	\begin{subfigure}[b]{0.24\textwidth}
		\centering
		\includegraphics[width=0.95\columnwidth]{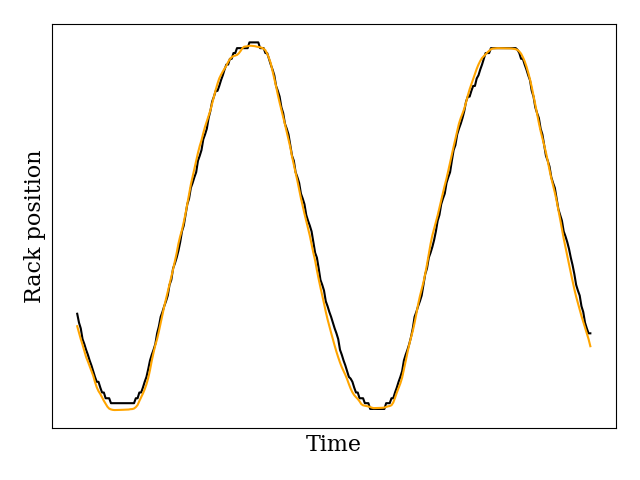}
		\caption{Data for sine excitation}\label{fig:sim_meas_data_rpc2}
	\end{subfigure}
	\caption{Data for rack position}\label{fig:sim_meas_data_rpc}
\end{figure}

\begin{figure}
	\centering
	\begin{subfigure}[b]{0.48\textwidth}
		\centering
		\includegraphics[width=0.95\columnwidth]{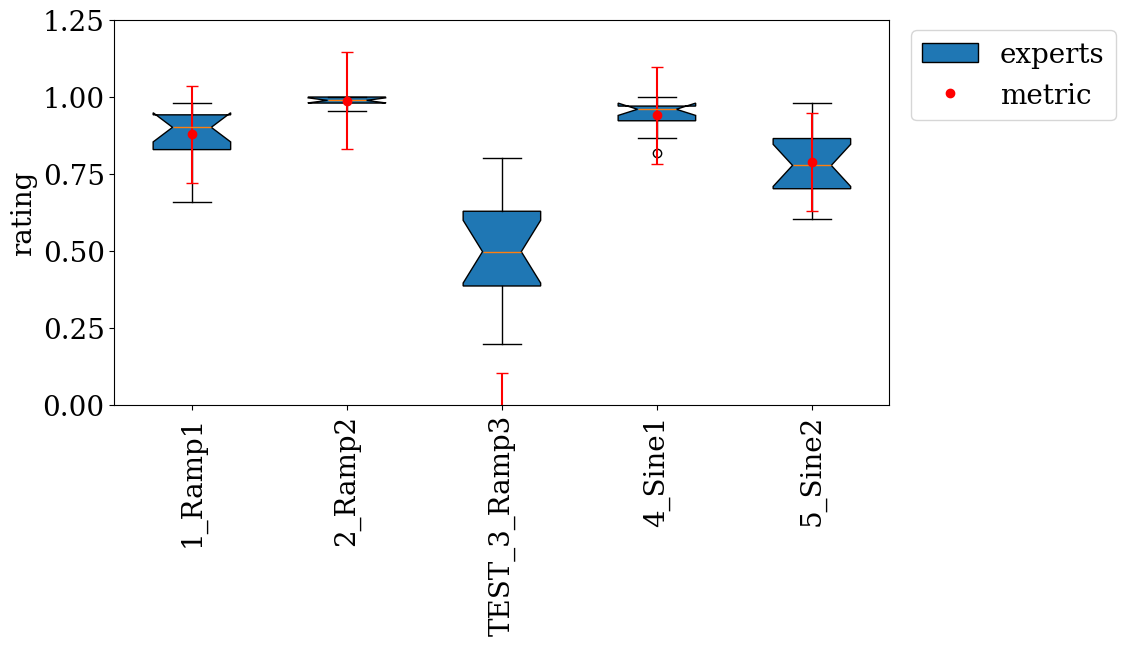}
		\caption{Simple prediction interval}\label{fig:metric_rack_1d}
	\end{subfigure}
	\hfill
	\begin{subfigure}[b]{0.48\textwidth}
		\centering
		\includegraphics[width=0.95\columnwidth]{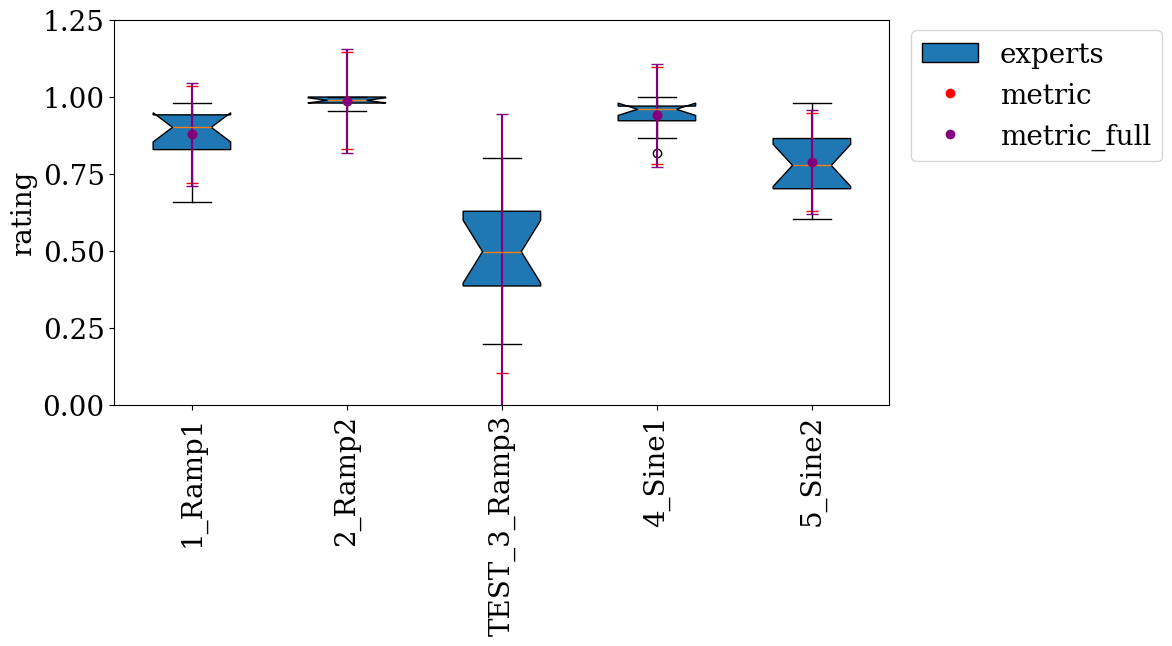}
		\caption{Full prediction interval}\label{fig:metric_rack_full}
	\end{subfigure}
	\caption{Metric result for rack position data}
\end{figure}

Lastly, we use the custom metric approach to rate the simulation results of a steering model. The steering system is shown in Fig.~\ref{fig:steering}. It is a steer-by-wire system, where an actatuator controls the driver-input dependent rack force on the steering rack. The position of the rack changes defines the direction in which the wheels than point.
The measurement and simulation data is given for different applied rack force signals. The output quantity of interest is the rack position. A set of rack position signals with a varying degree of agreement has been given to $14$ domain experts. Two examples of simulation and measurement data are shown in Fig.~\ref{fig:sim_meas_data_rpc}. The excitation signal is either of ramp-type or a sine wave.\\
According to section~\ref{sec:facevalidation}, the subject matter experts were given these time series plots and a sliding bar below the colorbar shown in Fig.~\ref{fig:ratingbar} as well as the explanation on the meaning of the rating given in Tab.~\ref{tab:expertrating}. No further background information on the measurements or the simulation was given.
Two of the 14 experts noted that for a more accurate rating of the measurement agreement further information on the testing device as well as the purpose of the simulation is needed. However, they where still able to give a rating. All other twelve experts proceeded with the rating without annotations. This may also reflect the different level of expertise that participating experts held. This partially explains the scattering of experts opinions shown in Fig.~\ref{fig:metric_rack_1d}.
The custom metric is fit to the data set and the remaining descriptive metrics in the resulting custom metric are mean absolute error, cross correlation as well as the combined error and the phase error of the Sprague-Geers metric. In Fig.~\ref{fig:metric_rack_full} becomes apparent, why the use of a confidence measure on the data is imperative to ensure the quality of the predictions of the custom metric. When using only the confidence measure on the training data as shown in red in Fig.~\ref{fig:metric_rack_1d}, the model predictions seem to match the training data quite well. However, as the yellow prediction interval in Fig.~\ref{fig:metric_rack_full} shows, the metric is not able to represent the experts opinion accurately on all data. Thus, a fit to the test and training data without a check for the remaining uncertainty would greatly overestimate the power of the prediction metric. In conclusion, more expert labeling is required to fit a precise custom metric in this use case.

\section{Conclusion and Outlook}\label{sec:conclusion}

We discussed why an accurate validation metric is use case dependent and general purpose metrics are often not sufficient. In contrast to the developed custom metric, a general purpose metric computes some rankable error from two curves but fails to define what level of error is still acceptable. Furthermore, some error types might be more relevant for one use case than the other. Based on that conclusion, we developed a systematic method to derive a custom metric based on subject matter experts. This metric closes the gap between error measure and a statement of the goodness of fit ('excellent', 'good', 'fair', 'poor'). Additionally, the specification of uncertainty induced to the custom metric by distributed subject matter opinions, available data quality and descriptiveness of the base metrics is crucial in validation. This is done by using prediction intervals.\\
The method has been demonstrated on three data sets: one artificial data set that allowed to study the influence of different hyperparameters, one data set taken from literature and one data set available to the authors. We showed that a good agreement of the custom metrics prediction and expert opinions can be reached. However, since this is a data driven approach, a sufficient amount of labeled data is necessary. For the literature example and our own use case only five labeled data points where available. This is not sufficient for usage in a real validation strategy. However, we could show that even for this low amount of data a fair prediction is possible. This is due to the linear regression ansatz, which is - compared to other data-driven approaches - rather robust for small data. As the collection of labeled data involves a lot of work, we recommend to stick with linear regression models. If for some cases large amount of data is available, more advanced methods like neural nets or decision trees might increase the accuracy of the custom metric further.

\section*{CRediT authorship contribution statement}
\textbf{Nicola Henkelmann:} Investigation, Methodology, Resources, Data Curation, Software, Visualization, Writing - Original Draft.
\textbf{Stephan Rhode:} Conceptualization, Methodology, Software, Validation,  Writing - Original Draft.
\textbf{Johannes von Keler:} Writing - Review \& Editing.

\bibliographystyle{IEEEtran}
\bibliography{IEEEabrv,references}

\end{document}